\documentclass[notitlepage,amsmath,amsfonts,amssymb,pra,reprint,floatfix]{revtex4-1}
\usepackage{graphicx}
\usepackage{bm}

\begin{document}
\title{Hartree-Fock symmetry breaking around conical intersections}

\author{Lena C. Jake}
\affiliation{Department of Chemistry, Rice University, Houston, TX 77005-1892}

\author{Thomas M. Henderson}
\affiliation{Department of Chemistry, Rice University, Houston, TX 77005-1892}
\affiliation{Department of Physics and Astronomy, Rice University, Houston, TX 77005-1892}

\author{Gustavo E. Scuseria}
\affiliation{Department of Chemistry, Rice University, Houston, TX 77005-1892}
\affiliation{Department of Physics and Astronomy, Rice University, Houston, TX 77005-1892}
\date{\today}

\begin{abstract}
We study the behavior of Hartree-Fock (HF) solutions in the vicinity of conical intersections. These are here understood as regions of a molecular potential energy surface characterized by degenerate or nearly-degenerate eigenfunctions with identical quantum numbers (point group, spin, and electron number). Accidental degeneracies between states with different quantum numbers are known to induce symmetry breaking in HF. The most common closed-shell restricted HF instability is related to singlet-triplet spin degeneracies that lead to collinear unrestricted HF (UHF) solutions. Adding geometric frustration to the mix usually results in noncollinear generalized HF (GHF) solutions, identified by orbitals that are linear combinations of up and down spins. Near conical intersections, we observe the appearance of coplanar GHF solutions that break all symmetries, including  complex conjugation and time-reversal, which do not carry good quantum numbers. We discuss several prototypical examples taken from the conical intersection literature. Additionally, we utilize a recently introduced a magnetization diagnostic to characterize these solutions, as well as a solution of a Jahn-Teller active geometry of H$_8^{+2}$.
\end{abstract}

\maketitle

\section{Introduction}
For the past several years we have been interested in the development of low-scaling computational methods for dealing with near-degenerate states, the so-called strong correlation problem \cite{YaoHessian, PQT, PHF}.  Accidental degeneracies, by which we mean degeneracies between states with different quantum numbers, are ubiquitous.  They occur, for example, in most molecular dissociations to open-shell fragments.  At the mean-field level, these degeneracies are associated with symmetry breaking, and can be detected by examining the eigenvalues of the symmetry adapted molecular orbital (MO) Hessian where they lead to well-studied instabilities of singlet, triplet and complex character \cite{p-benzyne, pople}.  Already, we have shown that symmetry projected methods in their variation after projection version \cite{GSmultiproj, ESfromPHF, 1DHubbardPHF} are capable of dealing with accidental degeneracies in a variety of practical contexts \cite{PHF_Esplitting, GHFfullerenes, 2DhubbardPHF}. 

The purpose of the present study is to extend our understanding of symmetry breaking and restoration to the situation of conical intersections (CX), where degenerate states share all of the same quantum numbers. These intersections or near intersections are necessary for any nonadiabatic process, providing a means of traveling between potential energy surfaces. They play an integral role in excited state dynamics and radiationless relaxation, explaining photochemical mechanisms for internal conversion.  Though they have been little explored, to use the words of Domcke, Yarkony, and K\"oppel, their presence seems to be ``the rule rather than the exception" in polyatomic molecules \cite{DomckeCXbook}.

It is convenient to discuss these intersections in terms of how the degeneracy is lifted. At a genuine CX, one can define two directions in which the degeneracy is lifted linearly, together forming what is known as the branching plane. These two directions are defined by the gradient difference (GD) and derivative coupling (DC) vectors, defined respectively as
\begin{subequations}
\begin{align}
\vec{x}_1 &= \frac{\partial (E_1-E_2)}{\partial \vec{R}},
\\
\vec{x}_2 &= \langle \psi_1 \frac{\partial \psi_2}{\partial \vec{R}} \rangle,
\end{align} 
\label{bpvecs}
\end{subequations}
where $E_1$ and $E_2$ are the energies of the intersecting states, $\psi_1$ and $\psi_2$ are their wave functions, and $\vec{R}$ are the nuclear coordinates.  The remaining $3N-8$ degrees of freedom will conserve the degeneracy, making up the  ``seam" of the two intersecting hypersurfaces. A consequence of this is that at least three atoms are necessary for such a crossing to occur \cite{HerzbergHiggins}. In addition to defining the branching plane, $\vec{x}_1$ can be used to optimize a conical intersection geometry, as it will point toward the apex of the cone when at a nearby geometry \cite{DomckeCXbook, theremustbeaCX_Robb,CXseam}.

To the best of our knowledge, the only application of symmetry projected Hartree-Fock (PHF) to conical intersections that has been carried out so far is in ozone \cite{ozone}, where non-orthogonal configuration interaction (NOCI) in the Hartree-Fock basis produces a qualitatively correct description of the relevant states. The limited number of degrees of freedom in ozone permits a scan of all degrees of freedom, a luxury not afforded by the molecules examined here. As a first step in this process, we explore the Hartree-Fock (HF) landscape in the branching planes of CXs optimized by the Complete Active Space Self Consistent Field (CASSCF) method.

The description of conical intersections at the Hartree-Fock level is complicated by the tendency of HF to break symmetries, which precludes assigning quantum numbers to states.  After all, if we cannot assign quantum numbers, we cannot meaningfully speak of degenerate states which have the same quantum numbers!  Symmetry projection, however, will restore these broken symmetries, and make the discussion of conical intersections meaningful again.  For now, we are interested simply in looking at which symmetries break and how those symmetries break in the vicinity of a CASSCF conical intersection.  Our motivation here is simple: our experience is that we should deliberately break and projectively restore those symmetries which might break spontaneously anyway.

The present work explores the HF landscape when all symmetries are allowed to break. Future investigations will examine the picture that emerges when they are restored. In the branching planes considered here, we observe intersections of unrestricted Hartree-Fock (UHF) excited states that occur near the CASSCF CX geometries. Complex coplanar generalized Hartree-Fock (GHF)  solutions are found in the vicinity of UHF degeneracies, in some cases interpolating between states of the same (conserved) quantum numbers. In our most detailed exploration, the branching plane of cyclobutadiene, we find multiple complex GHF intersections. To complement these examples of coplanar spin we present a noncoplanar GHF solution of tetrahedral H$_8^{+2}$.

Before we present our detailed results, however, let us take a moment to discuss the various kinds of Hartree-Fock solutions and how we might distinguish between them so we can properly decode what kinds of projection operators we will need in a symmetry restored treatment of conical intersections.

\section{Classes of Hartree-Fock Solutions}
The restricted Hartree-Fock (RHF) wave function is an eigenfunction of both $\hat{S}^2$ and $\hat{S}_z$ and is usually also an eigenfunction of time reversal ($\hat{\Theta}$), complex conjugation ($\hat{K}$), and point-group operators.  Unfortunately, RHF fails for strongly correlated systems, and one expects strong correlation in conical intersections as a matter of course due to the degeneracy.

Where RHF fails, one might use the unrestricted Hartree-Fock (UHF) formalism instead.  By permitting $\uparrow$- and $\downarrow$-spin electrons to occupy different spatial orbitals, UHF provides better energies at the cost of $\hat{S}^2$ symmetry. Typically UHF solutions also break point group symmetry; they must break at least one of complex conjugation and time-reversal symmetries.  Most UHF calculations result in real orbitals and are hence $\hat{K}$ eigenfunctions.

Sometimes UHF also breaks down, and one must allow for a generalized Hartree-Fock (GHF) approach \cite{GHFCarlos} in which $\hat{S}_z$ symmetry breaks in addition to $\hat{S}^2$.  By breaking $\hat{S}_z$ symmetry, GHF solutions provide noncollinear spin arrangements.  As with UHF, GHF solutions also usually break point group symmetry and must break at least one of time-reversal and complex conjugation symmetries.  Real GHF solutions which are $\hat{K}$ eigenfunctions have coplanar spin densities \cite{magtest} while complex GHF solutions may have coplanar or noncoplanar spin densities.  Noncoplanar spin has been seen previously in systems where high symmetry geometries induce spin frustration \cite{magtest, chromium} and in model Hamiltonians such as the Hubbard or Ising models  \cite{Resta_Berryreview, batista}.  
Though GHF significantly improves the shortcomings of RHF and UHF in strongly correlated systems, it has not been regularly used in the community. A complete table classifying HF solutions by the symmetries they preserve can be found in both \cite{GHFCarlos} and \cite{magtest}, though note that this classification was first carried out by Fukutome \cite{Fukutome_UHFapps} and has also been discussed extensively by Stuber \cite{Stuber_Paldus_SymmbreakingIPM}.  

As we have alluded to earlier, one can check whether there is a lower-energy HF solution by considering the eigenvalues of the MO Hessian.  A negative eigenvalue indicates the presence of a more stable solution in the direction of the corresponding eigenvector.  By taking particular blocks of the Hessian, one can limit one's testing to consider instabilities to a particular symmetry block \cite{pople}.  For example, one could test whether an RHF solution is unstable toward other RHF solutions, or toward UHF solutions, i.e. one can look for singlet or triplet instabilities.  Similarly, one can test whether UHF solutions are unstable toward GHF wave functions, and one can test for instabilities toward solutions which break complex conjugation symmetry. 

Note that a degeneracy between occupied and virtual orbitals guarantees a negative diagonal element in the MO Hessian and therefore an instability, but is not necessary for such an instability to exist \cite{m0m1instabilities_Yamada}. Symmetry broken solutions in the branching planes explored here provide further counterexamples.  Other cases include the noncollinear solutions found in fullerene molecules \cite{GHFfullerenes,magtest}, where large band gaps persist as RHF succumbs to UHF and ultimately GHF.

The number of zero eigenvalues can also yield information regarding symmetry breaking and stability, as a UHF or GHF solution will acquire improper zero modes as an artifact of symmetry breaking.  The simplest example of this would be the dissociation of H$_2$, where beyond the Coulson-Fischer point UHF yields a more stable solution than the singlet RHF that is stable at equilibrium bond length. Past this point, the lowest Hessian eigenvalue of the RHF solution becomes negative, while the UHF solution has two improper zero modes due to breaking $\hat{S}_x$ and $\hat{S}_y$ \cite{YaoHessian}. For an equally simple example of a triplet instability, the interested reader might examine HF solutions to the Be atom \cite{GHFCarlos, YaoHessian}.

\begin{figure*}[ht]
    \includegraphics[width=0.25\textwidth]{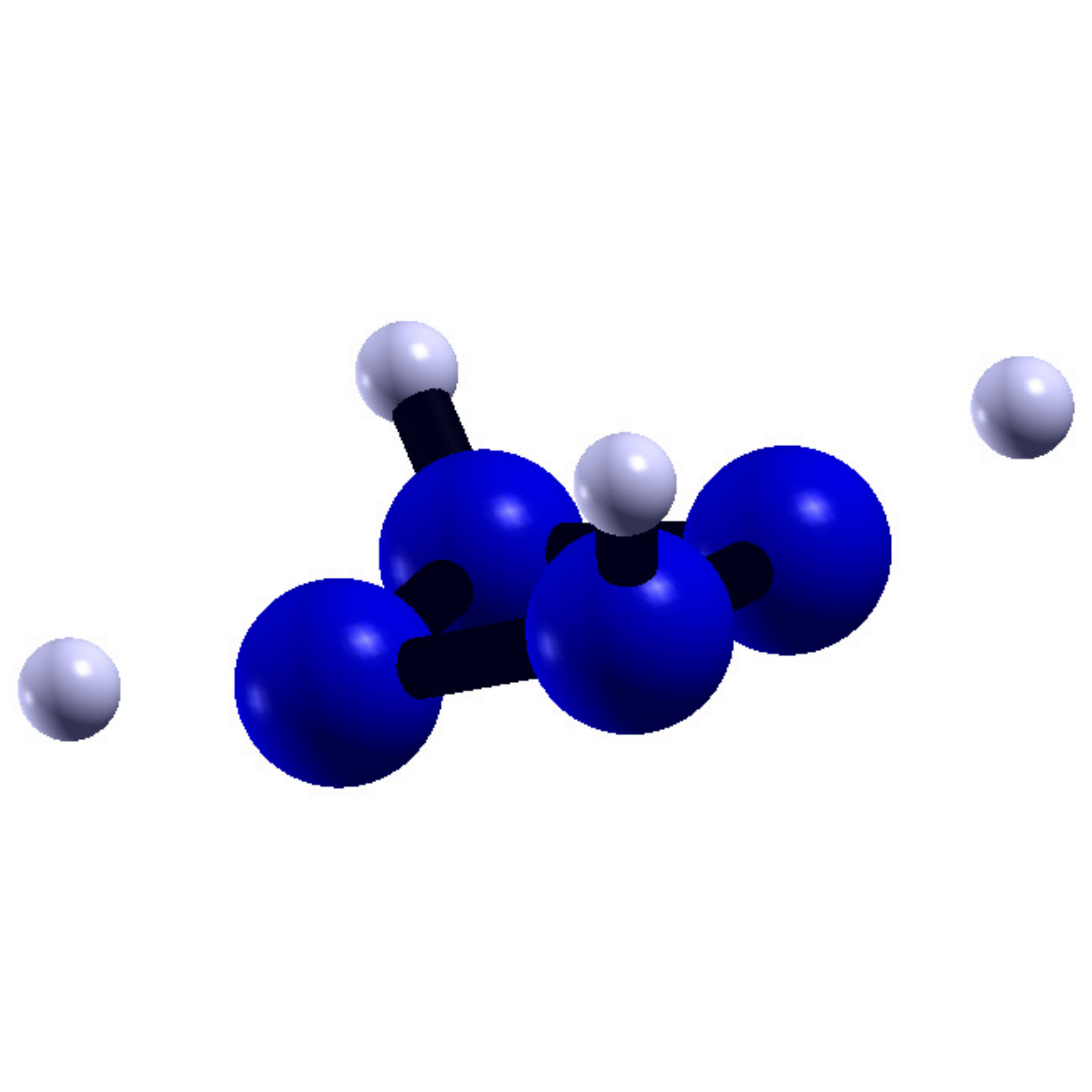} \hfill 
    \includegraphics[width=0.25\textwidth]{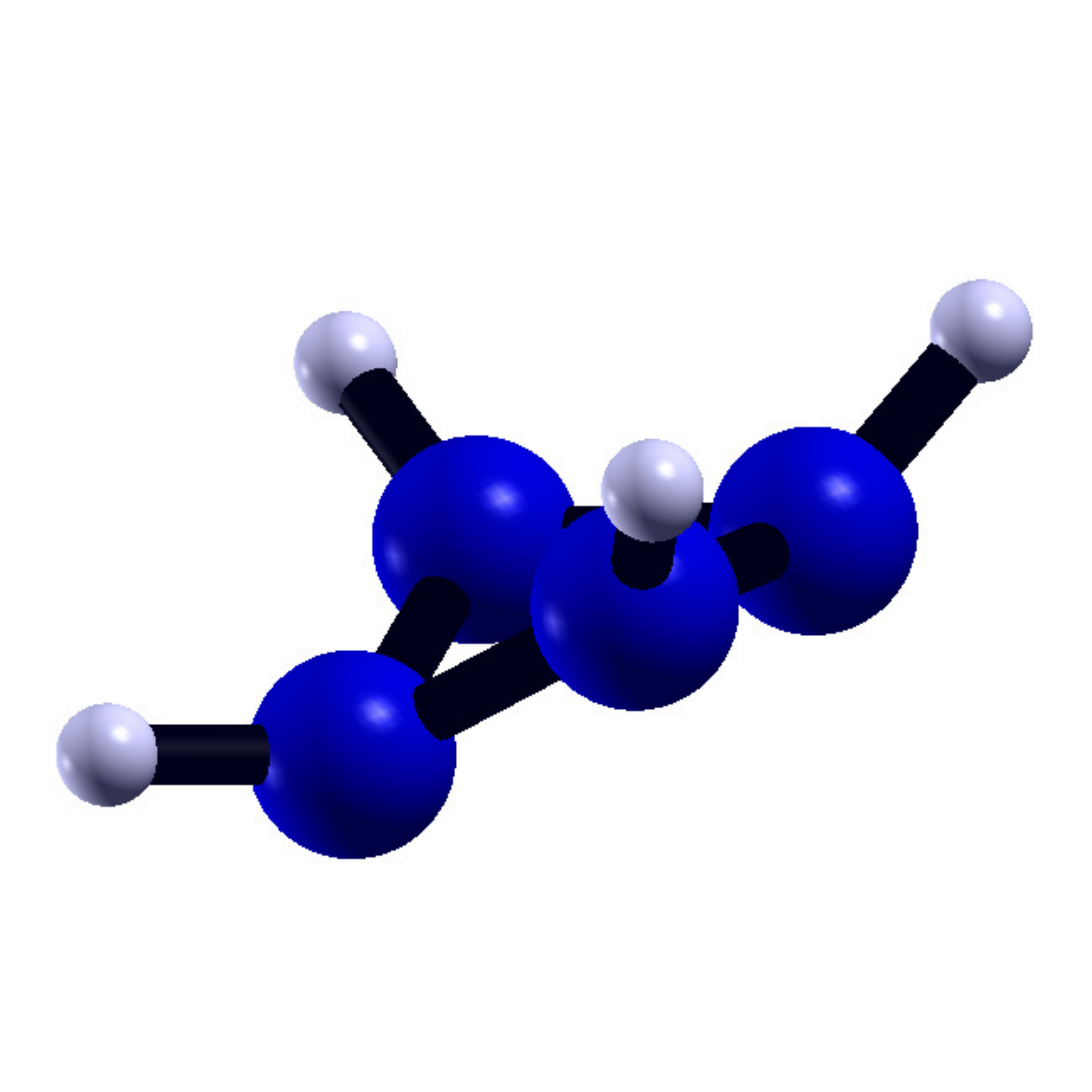} \hfill 
    \includegraphics[width=0.25\textwidth]{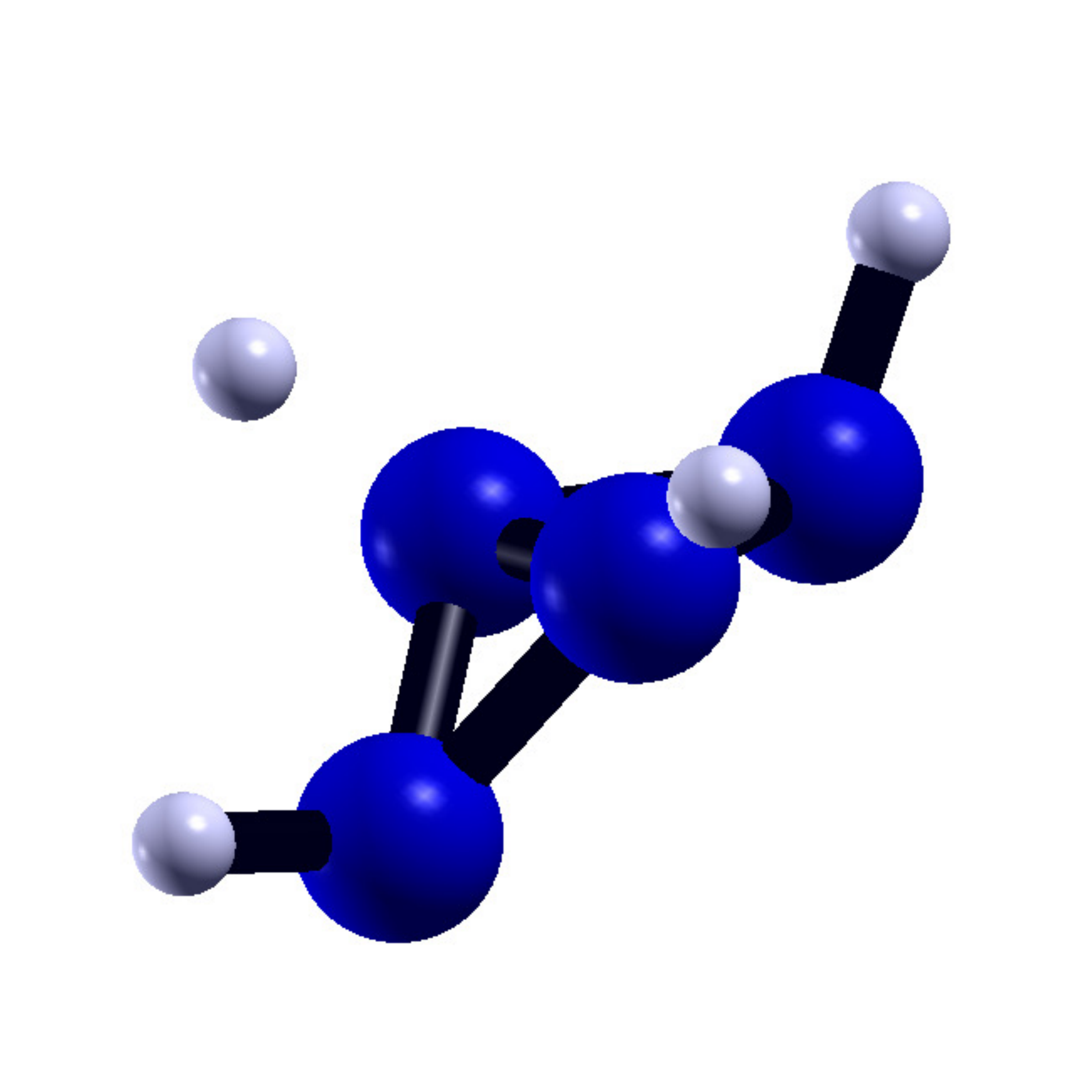} \\
    \includegraphics[width=0.25\textwidth]{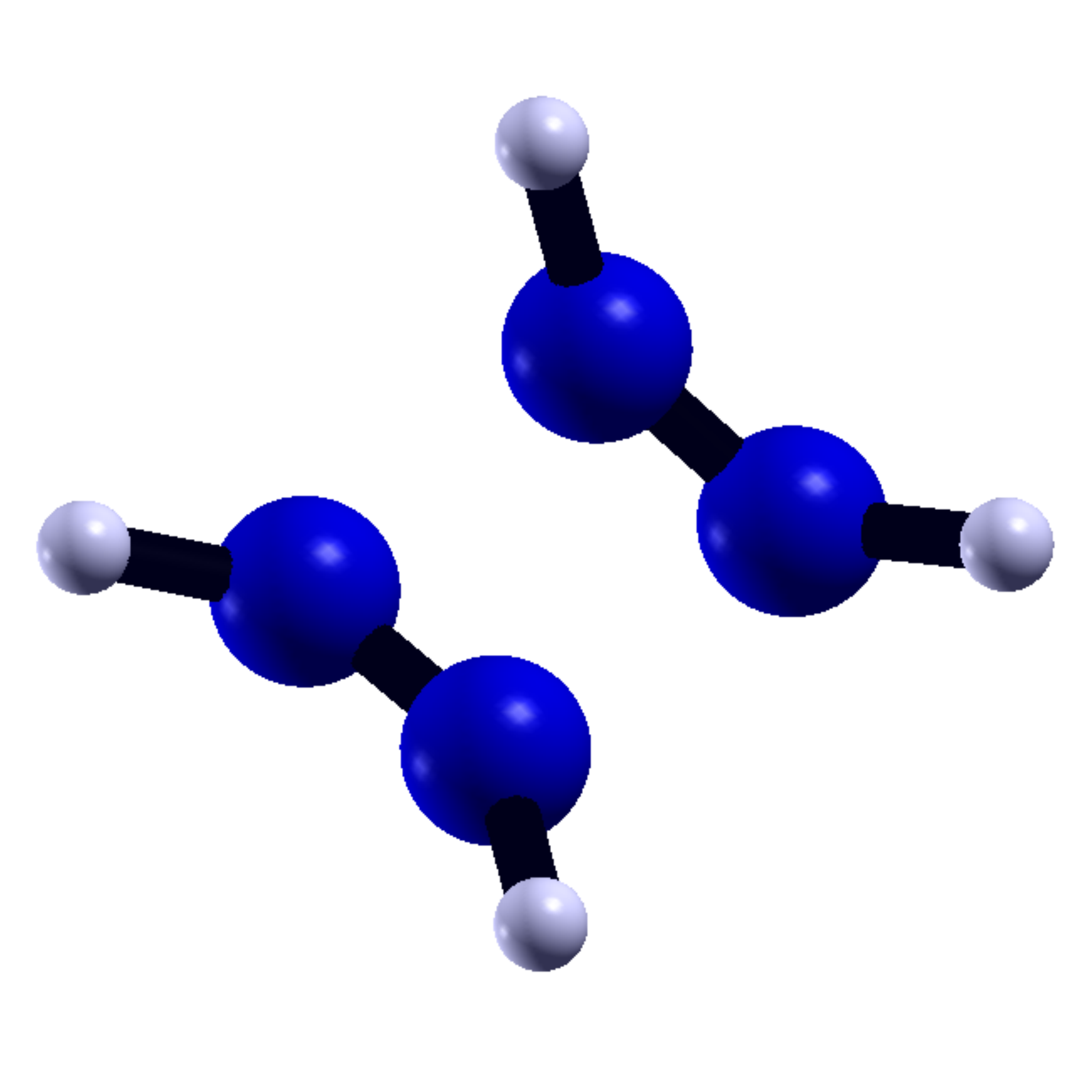} \hfill 
    \includegraphics[width=0.25\textwidth]{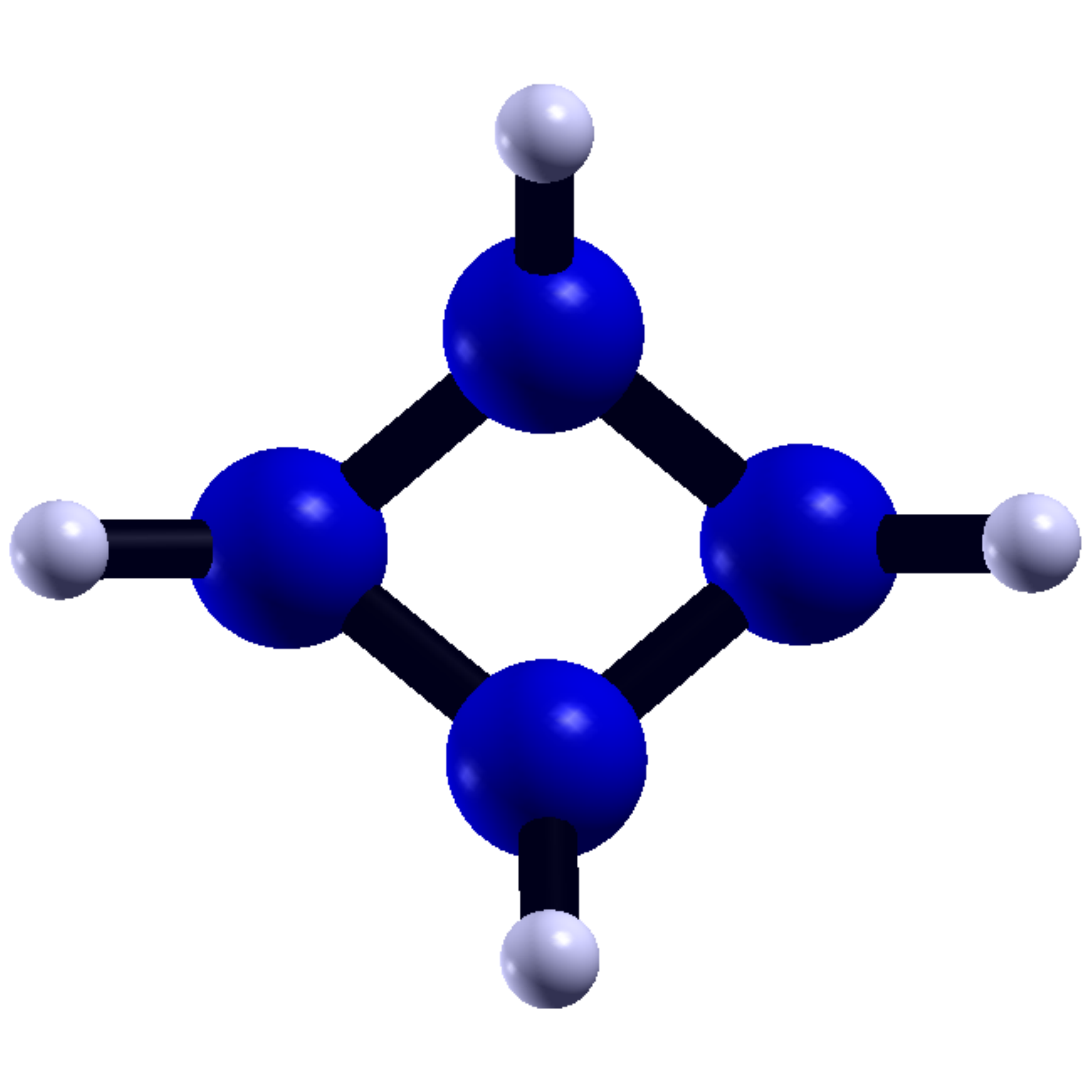} \hfill 
    \includegraphics[width=0.25\textwidth]{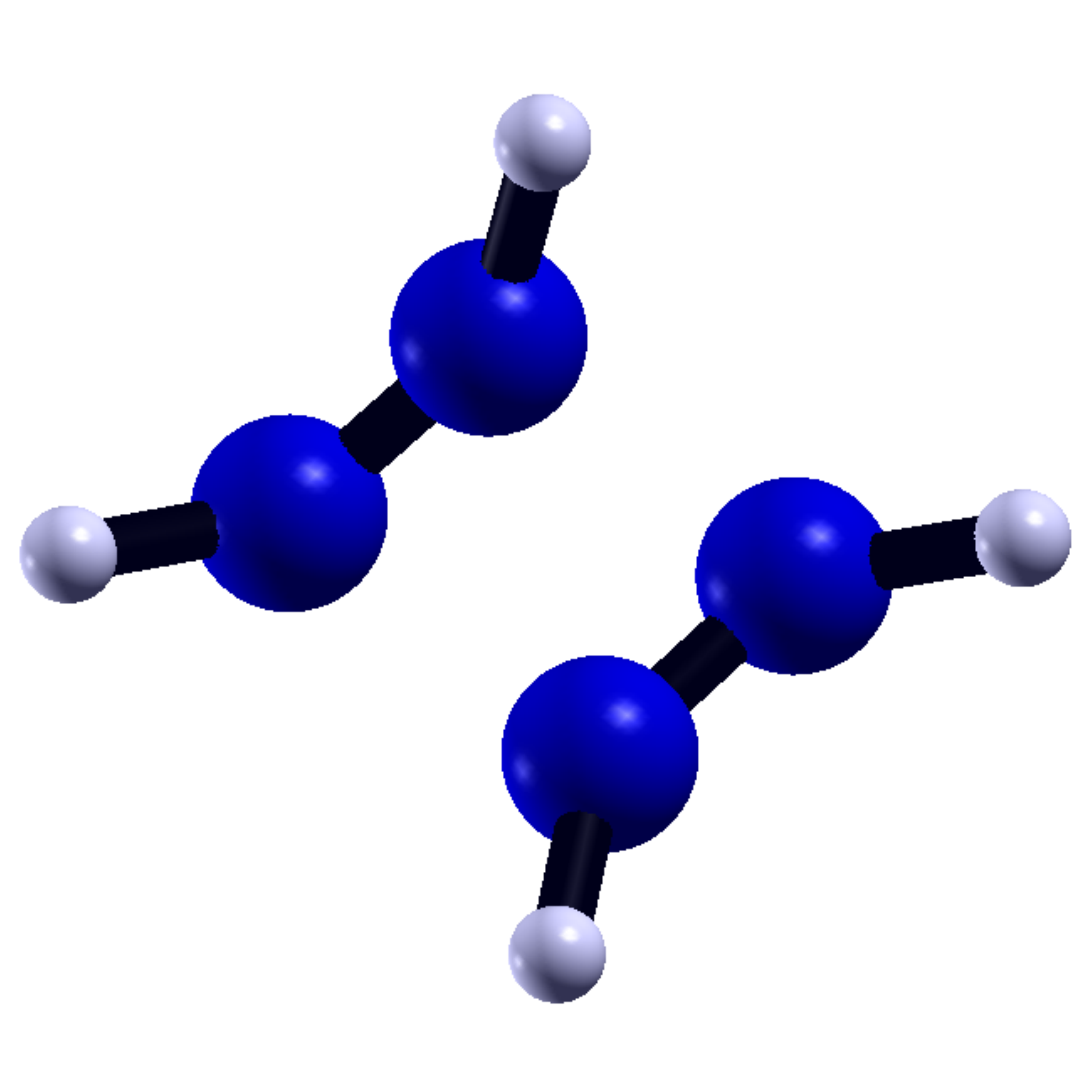}
\caption{The cyclobutadiene CX geometry (center, top and bottom) and CX displaced by $\vec{x}_1$ (top) and $\vec{x}_2$ (bottom) with weights of $\pm2$ (left and right).\label{fig:cbd_cx}}
\end{figure*}

\section{Determining Coplanarity}
It should be noted that while we have discussed the symmetry breaking permitted by different incarnations of HF, permitting a symmetry to break does not guarantee that it will. That is to say, a GHF search may still arrive at a UHF solution. In such a case, the UHF solution may even be an eigenfunction of spin in some other direction -- say, $\hat{S}_z$ -- and it is not immediately obvious how to tell a noncollinear solution from a rotated UHF solution.  Similarly, it is not necessarily simple to tell whether a GHF solution is coplanar or noncoplanar.

In the last few years, means of differentiating between collinear and noncollinear solutions have emerged. Small, Sundstrom, and Head-Gordon define a test \cite{colltest} which uses the fact that if a wave function is an eigenfunction of $\hat{S}_{\hat{n}}$ for some direction $\hat{n}$, then $\langle \hat{S}_{\hat{n}}^2 \rangle - \langle \hat{S}_{\hat{n}} \rangle^2 = 0$.  Thus, if the matrix $\langle \hat{S}_i\hat{S}_j \rangle - \langle \hat{S}_i \rangle \langle \hat{S}_j \rangle$ has any zero eigenvalues, the solution must be collinear. This revelation is integral to the diagnostic used here, but in its original formulation has the drawback of relying on the two-particle density matrix. In a recent paper \cite{magtest}, we have shown a simplified test which is identical to that of Small and coworkers for single determinants and which can discriminate between coplanar and noncoplanar solutions.

\begin{table}[t]
\caption{Characterization of HF Solutions
\label{Tab:Tzeroevals}}
\begin{tabular}{ccl}
\hline\hline
\# Zero Eigenvalues &
\# Zero Eigenvalues & 
\\
of $\mathbf{T}$  &
of $\bm{\tau}$   &
Characterization
\\
\hline
3    &  3    & nonmagnetic (RHF)
\\
2    &  2-3  & collinear (UHF)
\\
0-1  &  1-3  & coplanar (GHF)
\\
0    &  0    & noncoplanar (GHF)
\\
\hline\hline
\end{tabular}
\end{table}

The density matrix $\bm{\gamma}$ can be decomposed into its $x$, $y$, and $z$ spin components as
\begin{subequations}
\begin{align}
\mathbf{M}_x &= \bm{\gamma}_{\uparrow\downarrow} + \bm{\gamma}_{\downarrow\uparrow},
\\
\mathbf{M}_y &= \mathrm{i} \, (\bm{\gamma}_{\uparrow\downarrow} - \bm{\gamma}_{\downarrow\uparrow}),
\\
\mathbf{M}_z &= \bm{\gamma}_{\uparrow\uparrow} - \bm{\gamma}_{\downarrow\downarrow}.
\end{align}
\label{decompden}
\end{subequations}
If spin rotations can make two of these components vanish, the density matrix is collinear.  If spin rotations can make one of these components vanish, the density matrix is coplanar.  We can check this possibility by diagonalizing the matrix $\mathbf{T}$ with components
\begin{equation}
\mathrm{T}_{ij} = \mathrm{Tr}(\mathbf{M}_i \, \mathbf{S} \, \mathbf{M}_j \, \mathbf{S})
\label{colltest}
\end{equation}
where $\mathbf{S}$ is the overlap matrix.  Collinear determinants correspond to one non-zero eigenvalue of $\mathbf{T}$, while for noncollinear determinants $\mathbf{T}$ has two or three non-zero eigenvalues.

\begin{figure*}[t]
\includegraphics[width=0.48\textwidth]{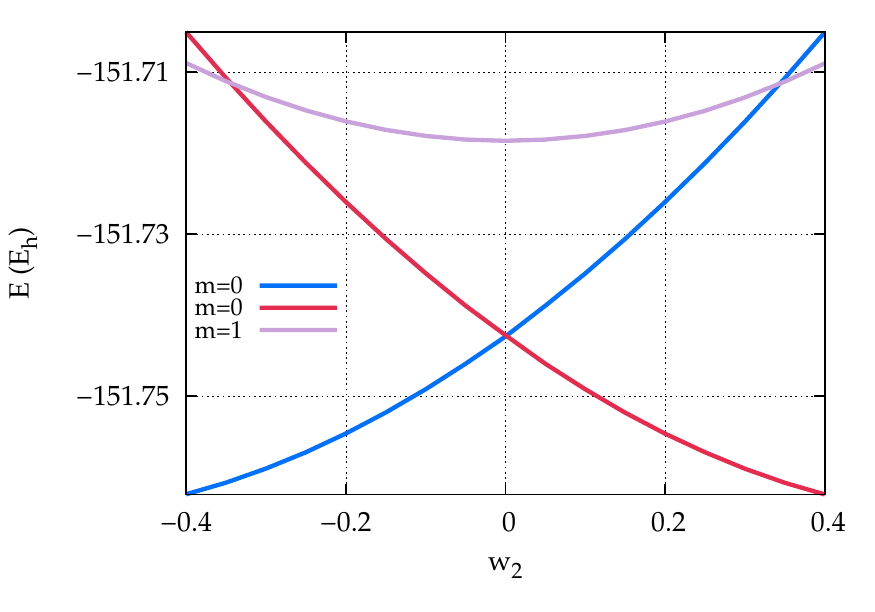}
\hfill
\includegraphics[width=0.48\textwidth]{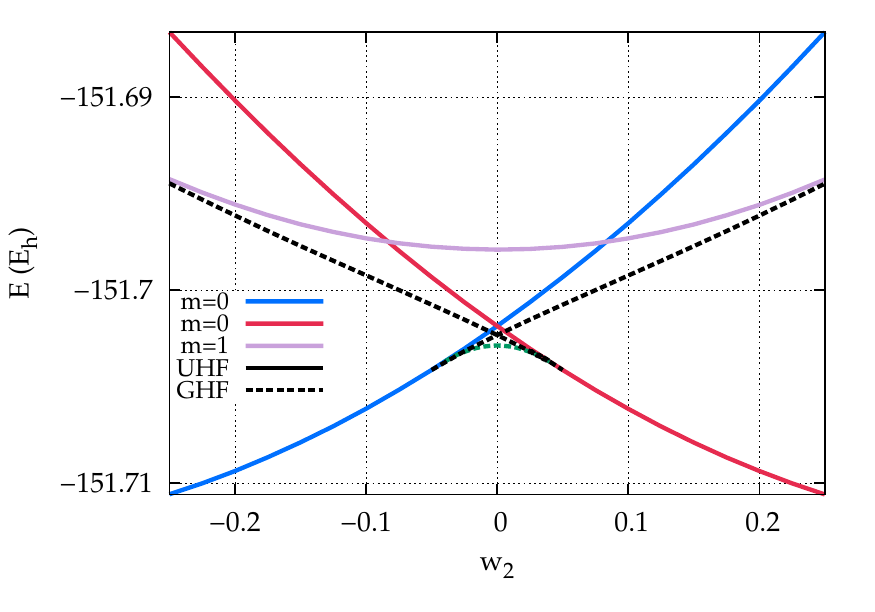}
\caption{
Energies for displacement of the cyclobutadiene CX by the DC vector $\vec{x}_2$, plotted as a function of displacement weight $w_2$.  Left panel: CASSCF(4,4) energies. Right panel: HF energies, with regions of noncollinear spin indicated by dashed lines. Collinear solutions are labeled by the $m$ quantum number associated with $\hat{S}_z$. Not pictured is the stable HF solution, which is collinear with $m=0$.
\label{fig:cbd_energies}}
\end{figure*}

While this test is identical to that of Small and coworkers for single determinants, we can generalize it slightly to test for coplanarity.  Noncoplanar density matrices mean that all three of $\mathbf{M}_x$, $\mathbf{M}_y$, and $\mathbf{M}_z$ must have non-zero real parts.  Thus, we can distinguish coplanar from noncoplanar GHF solutions by diagonalizing the related matrix $\bm{\tau}$ with components
\begin{equation}
\tau_{ij} = \mathrm{Tr}[Re(\mathbf{M}_i) \, \mathbf{S} \, Re(\mathbf{M}_j) \, \mathbf{S}].
\label{coptest}
\end{equation}
For coplanar GHF solutions, $\bm{\tau}$ has a zero eigenvalue.  Our test is summarized in Table \ref{Tab:Tzeroevals}.  Note that we order eigenvalues so that after rotation, $\mathrm{T}_{zz} \ge \mathrm{T}_{xx} \ge \mathrm{T}_{yy}$, and similarly for eigenvalues of $\bm{\tau}$. More details about this magnetization diagnostic can be found in \cite{magtest}.

\section{Results}
In its most recent version, the \textit{Gaussian} suite of programs only supports CX optimization using the CASSCF method. The resulting geometry and branching plane are not necessarily the same as those defined by a PHF degeneracy, and it is not guaranteed that such a degeneracy could be classified as a CX at all. In this work, we follow HF solutions in the CASSCF branching plane.

Conical intersection geometries were optimized using equally-weighted state-averaged CASSCF calculations with no symmetry constraints, as implemented in \textit{Gaussian16} \cite{g16}.  The more affordable spin-free Hartree-Waller determinants were used in CASSCF calculations, and therefore where singlets and triplets are shown together it should be noted they come from separate state-averaged calculations. Active spaces were defined as only the $\pi$ orbitals and electrons, and all calculations, CASSCF or HF, were carried out in the STO-3G basis. This minimal basis set was used in an effort to avoid smearing static correlation effects with those of dynamic correlation.  The CXs of aromatic and antiaromatic molecules are well documented in computational organic chemistry literature \cite{cbdCX,benzeneCX,DomckeCXbook,styreneCX,fulveneCX}, providing starting points for geometry optimizations.

\begin{figure}[t]
\includegraphics[width=0.48\textwidth]{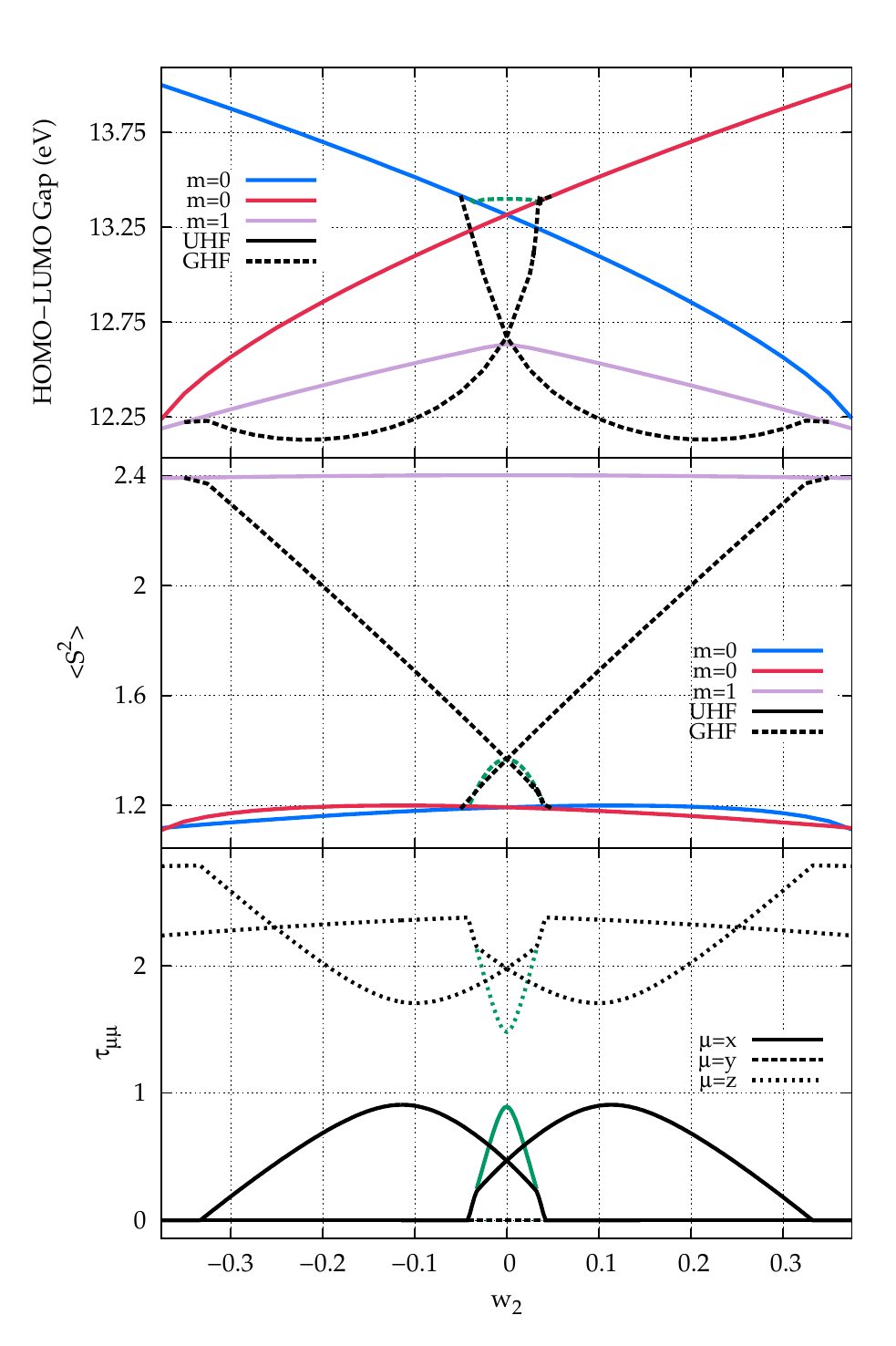}
\caption{
HF band gap, $\langle \hat{S}\rangle$, and eigenvalues of $\mathbf{\tau}$ (Eq. \ref{coptest}) for displacement of the cyclobutadiene CX by the DC vector $\vec{x}_2$, plotted as a function of displacement weight $w_2$. Line style and color scheme are consistent with Fig. \ref{fig:cbd_energies}.
\label{fig:cbd_hf}}
\end{figure}

A variety of initial guesses yielded many UHF solutions, which were in turn followed as the geometry was displaced by a range of weights of the branching plane vectors. Where we refer to singlet or triplet UHF solutions, we should clarify that this is in reference to the $m$ quantum number associated with $\hat{S}_z$, rather than the $s$ quantum number associated with $\hat{S}^2$.  To find GHF solutions, we destroyed $\hat{S}_z$ symmetry with application of a Fermi-Contact perturbation to the converged UHF and halted the resulting calculation after several iterations. This symmetry broken initial guess served as a starting point for a GHF calculation with no perturbation.  Directions for the perturbation were selected from linear combinations of branching plane vectors, with the motivation that the directions that lift degeneracy should also make convergence to a new, symmetry broken solution more likely than convergence back to either of the intersecting collinear surfaces.

A modified development version of \textit{Gaussian} \cite{gdv-h21} carried out the collinearity test of Small and coworkers. Once a solution was identified as noncollinear, in-house code was used to determine coplanarity by diagonalization of $\bm{\tau}$ of Eq. \ref{coptest}. Another modified development version of \textit{Gaussian} \cite{gdv-h21} calculated the GHF Hessian to determine stability. The number of Hessian zero modes, in addition to reflecting the symmetry breaking we show with the magnetization diagnostic, will also be used to detect degeneracies, near degeneracies, or symmetric invariances. Where we present molecular geometries, these figures have been created using the X-Window Crystalline Structures and Densities software \cite{xcrysden}. Below, we discuss HF in the branching plane of four different CXs, with a focus on that of cyclobutadiene. We observe intersecting UHF states near the CASSCF CX in all cases, and in cyclobutadiene we see complex coplanar GHF solutions cross as well. In each branching plane we converged to complex coplanar GHF for geometries around UHF intersections.

\subsection{Cyclobutadiene}
A CX was optimized between the first two singlets of cyclobutadiene, resulting in a loosely defined $C_s$ geometry (Fig. \ref{fig:cbd_cx}) where the ring has been bent to a 25$^\circ$ degree dihedral angle and the hydrogen atoms pulled out of plane \cite{cbdCX}. At this geometry, the CASSCF energy difference between these two states is 0.2 kcal/mol. 

Motion along the DC vector $\vec{x}_2$ corresponds to shortening and lengthening of alternate C-C bonds, resulting in dissociation into different $\textrm{C}_2\textrm{H}_2$ fragments in the positive and negative directions (Fig. \ref{fig:cbd_cx}). Along $\vec{x}_2$, the CASSCF excitation energy remains linear until weights of about $\pm$1.25, and the intersecting states are symmetric about the CX (Fig. \ref{fig:cbd_energies}). Along the GD vector $\vec{x}_1$, displacement from the CX geometry results in a more bent dihedral angle in the four carbons, and an increase in alternating bond angles of the ring (Fig. \ref{fig:cbd_cx}). Motions in the positive and negative directions have less symmetric effects on the CASSCF energy than seen along the DC vector, and the excitation energy becomes nonlinear before weights of $\pm$0.5. 

As the CX geometry is displaced along $\vec{x}_2$ (Fig. \ref{fig:cbd_energies}), two singlet UHF states intersect at a geometry very near the CASSCF CX, with a triplet UHF solution about 2 kcal/mol above them. Complex coplanar GHF solutions interpolate between each of the singlet states and the triplet, intersecting 0.3 kcal/mol below the UHF singlets. At the point of intersection, other properties of the GHF solutions coincide as well (Fig. \ref{fig:cbd_hf}). Another coplanar GHF, not pictured in Fig. \ref{fig:cbd_energies} or \ref{fig:cbd_hf}, branches off of the UHF triplet and vanishes before rejoining the UHF singlet. A fourth complex coplanar GHF solution, plotted in green, exists only for a small range of $w_2$ around the CX geometry, another 0.3 kcal/mol below the GHF intersection. While this appears at first glance to interpolate between the singlet UHF states, it vanishes before reaching either. In all cases where these GHF disappear, spin properties change dramatically as this geometry is approached, seeming to signal the convergence failure to come.  None of the Hartree-Fock states described thus far is the ground state; the stable solution is collinear with $m=0$ and 16 kcal/mol lower, giving all states at least one negative Hessian eigenvalue.

\begin{figure}[t]
\includegraphics[width=0.48\textwidth]{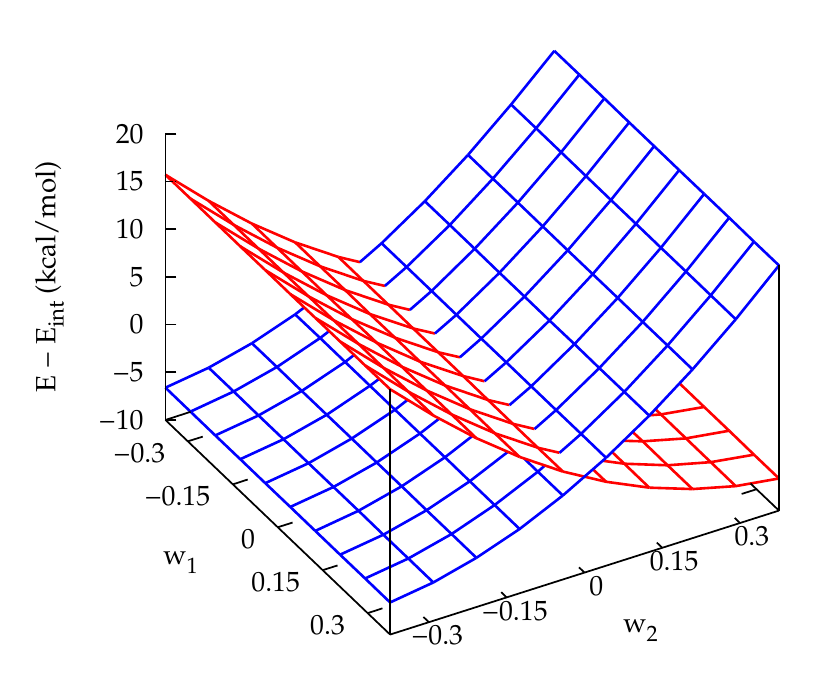}
\caption{
Energies of two intersecting UHF solutions in the branching plane of cyclobutadiene, plotted as the difference from the energy at their intersection. Color scheme is consistent with that of Fig. \ref{fig:cbd_energies} and \ref{fig:cbd_hf}.
\label{fig:cbd_3dbp}}
\end{figure}

\begin{figure*}[t]
    \includegraphics[width=0.25\textwidth]{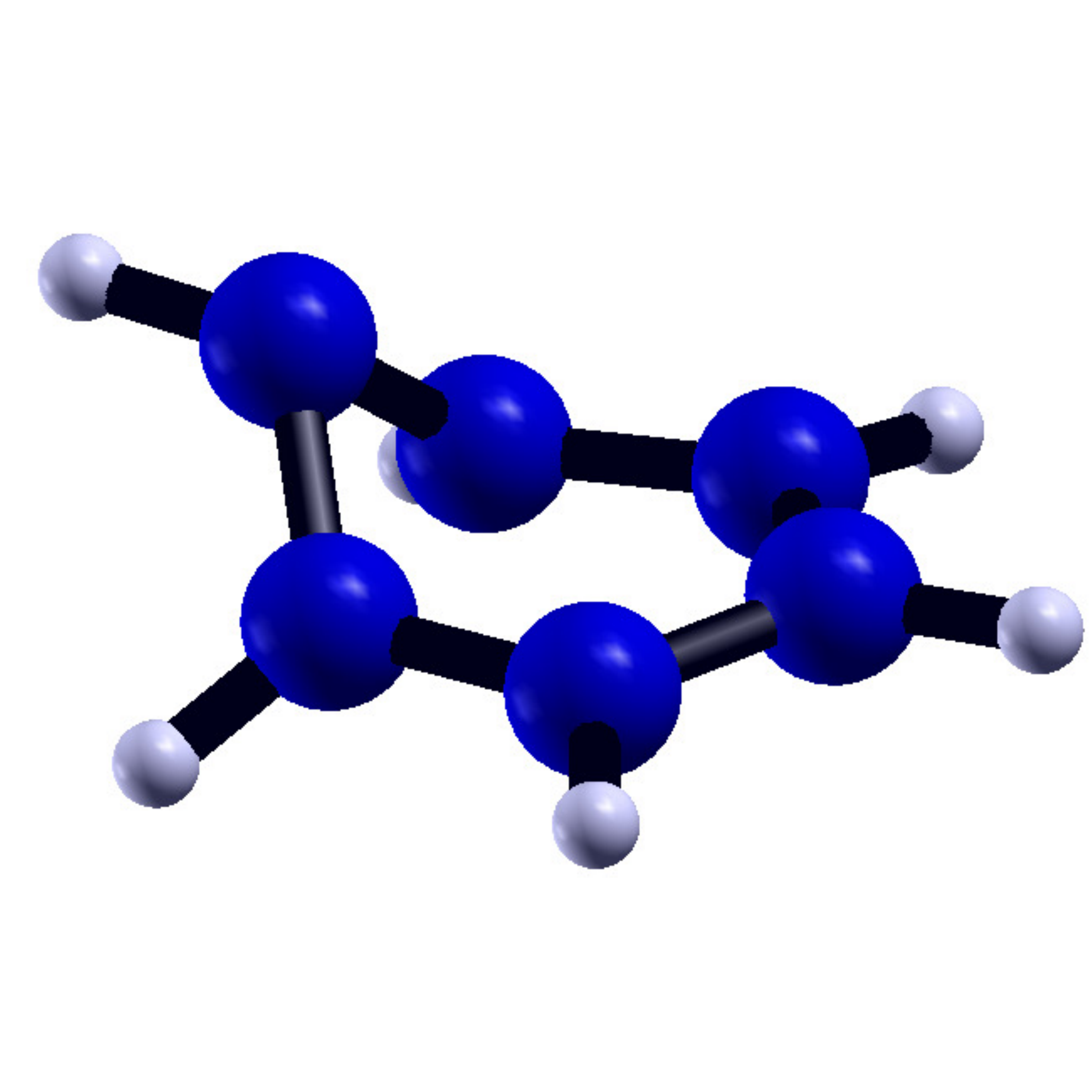} \hfill
    \includegraphics[width=0.25\textwidth]{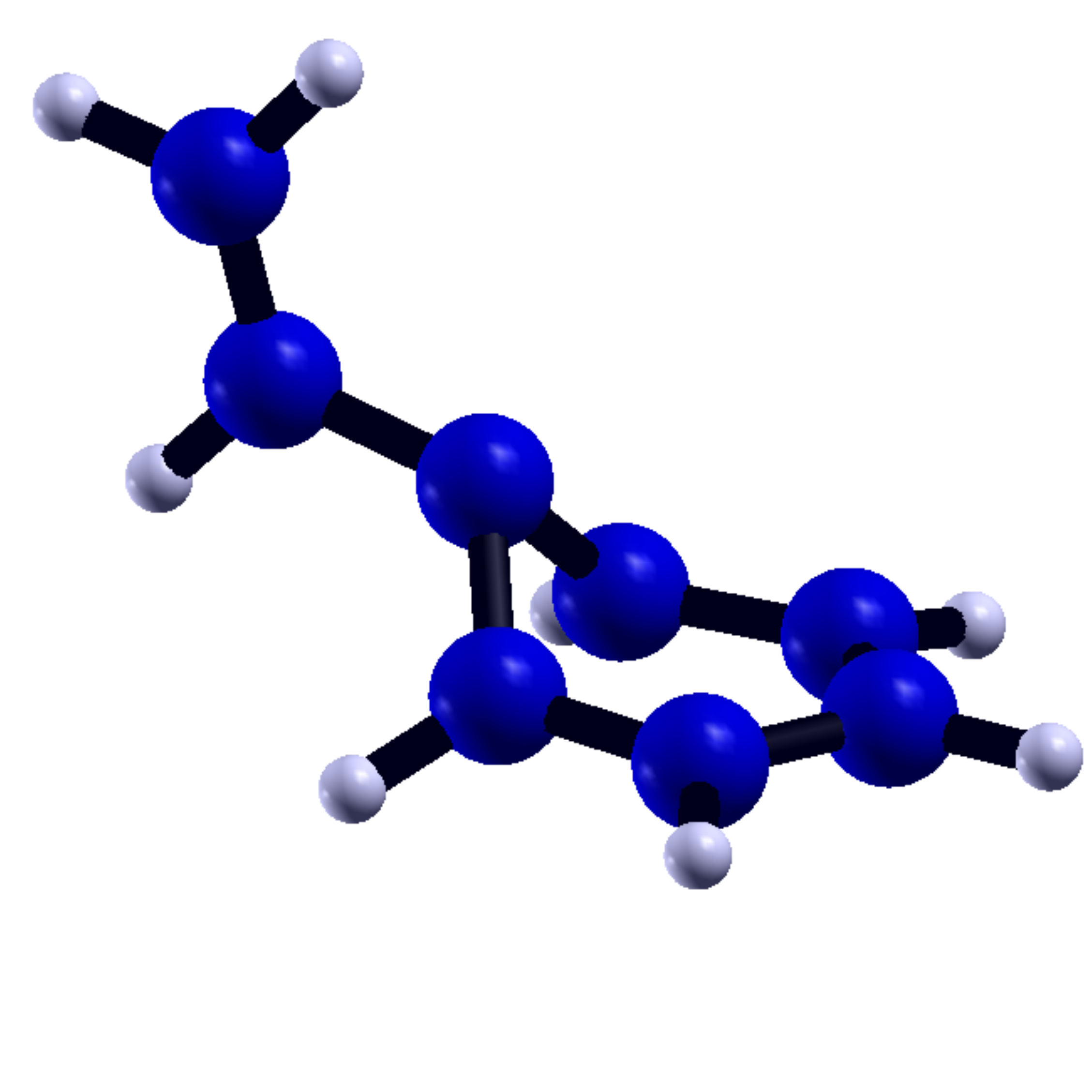} \hfill
    \includegraphics[width=0.25\textwidth]{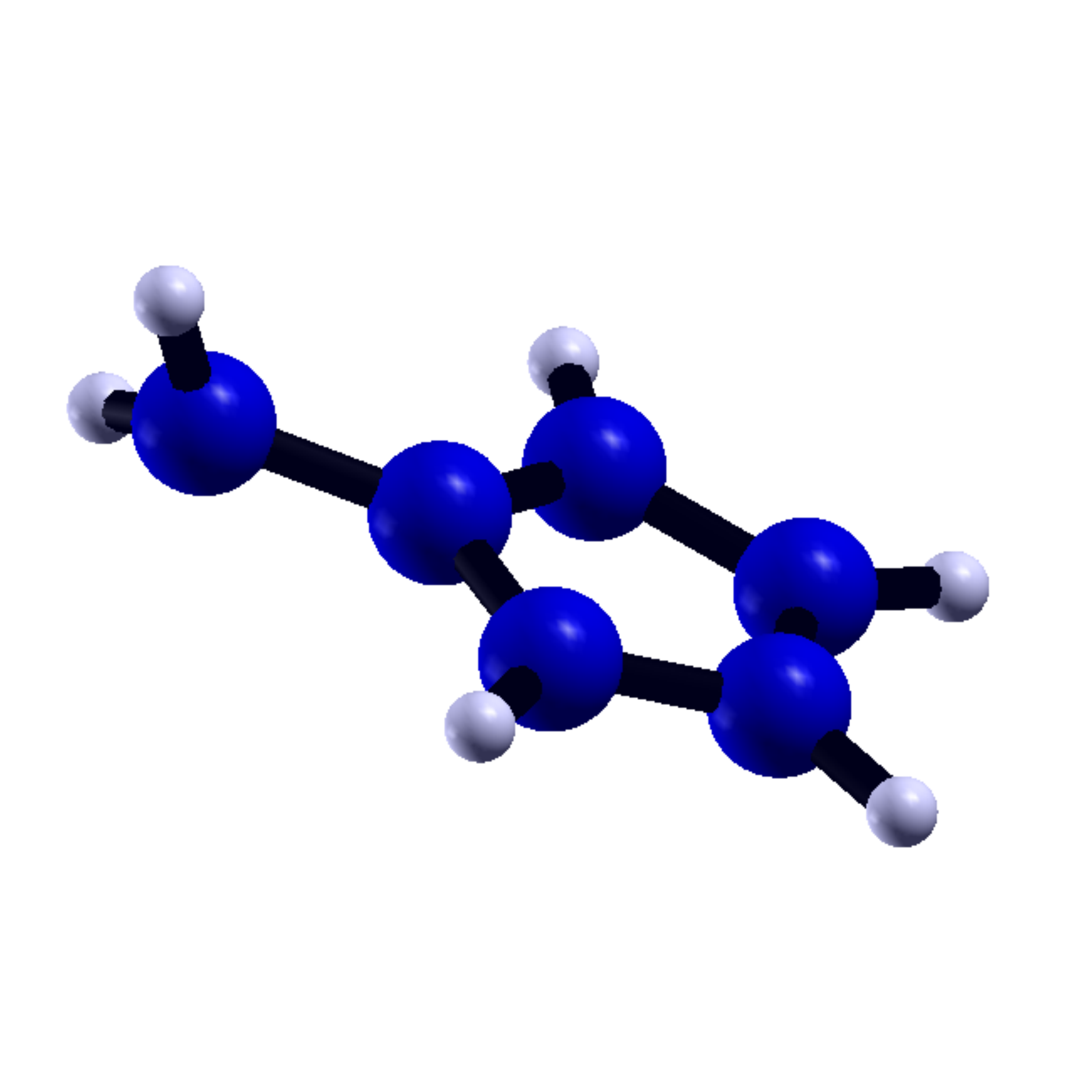} \hfill
\caption{Left to right: The benzene, styrene, and fulvene CX geometries.\label{fig:other3cx}}
\end{figure*}

It is worth noting that for a range of weights the lowest energy GHF has four Hessian zero modes. Three can be attributed to symmetry breaking, while the fourth indicates a quasi-symmetric invariance that we have not been able to fully identify. This unaccounted for zero-mode emphasizes the need for future work investigating HF around this CX. Another interesting trait of this GHF solution is its MO structure. For each of the UHF solutions, orbital energies occur in degenerate pairs around the CX. While the intersecting GHF solutions do not reflect the loose $C_s$ symmetry of the CX geometry in the same way, this lowest energy GHF does.

While the CASSCF degeneracy is lifted along the GD vector, both the UHF and GHF solutions seen intersecting along $\vec{x}_2$ remain nearly degenerate as they are  followed along $\vec{x}_1$ (Fig.  \ref{fig:cbd_3dbp}), a clear deviation from the branching plane behavior we would expect.  Rather than restoring $\hat{S}_z$ symmetry for negative $w_1$, the GHF solutions remain noncoplanar. The GHF solutions that vanish along $\vec{x}_2$ also vanish for negative displacements along $\vec{x}_1$, at a weight of just over $w_1=-0.25$. All GHF solutions continue in the positive direction and restore collinearity before $w_1=0.5$. It seems that while degeneracies of HF states occur very near the CASSCF CX geometry, the motions corresponding to the CASSCF branching plane vectors do not lift the degeneracies of HF states in quite the same way.  This suggests that the CASSCF and projected Hartree-Fock branching planes will be distinct.

\begin{figure}[b]
    \includegraphics[width=0.48\textwidth]{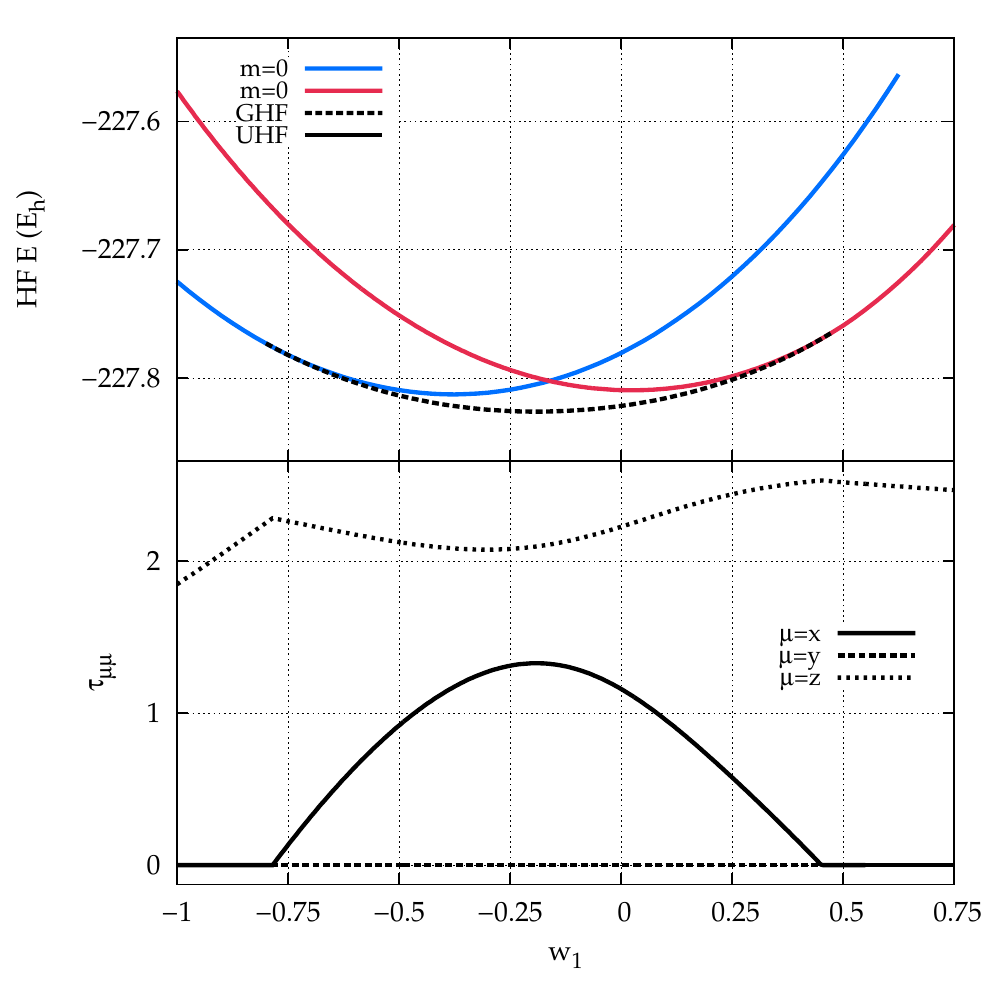}
\caption{HF energy and eigenvalues of $\mathbf{\tau}$ (Eq. \ref{coptest}), for fulvene along the GD vector $\vec{x}_1$, plotted as a function of displacement weight $w_1$.\label{fig:fulvene}}
\end{figure}

\subsection{Benzene, Fulvene, and Styrene}
CXs between singlets in benzene\cite{benzeneCX,CXinspec_Yarkony} and styrene\cite{CXinspec_Yarkony,styreneCX} share similar geometries with loose $C_s$ symmetry, such that one atom of the ring is pushed out of plane to a pre-fulvene-like puckered ring (Fig. \ref{fig:other3cx}). For each, displacement along the DC vector $\vec{x}_2$ results in pushing the out of plane moiety either further out of or into the plane of the ring, depending on the direction of displacement. Neither CX is the ground state, each being less than 30 kcal/mol above the stable CASSCF triplet.

\begin{figure*}[t]
    \includegraphics[width=0.48\textwidth]{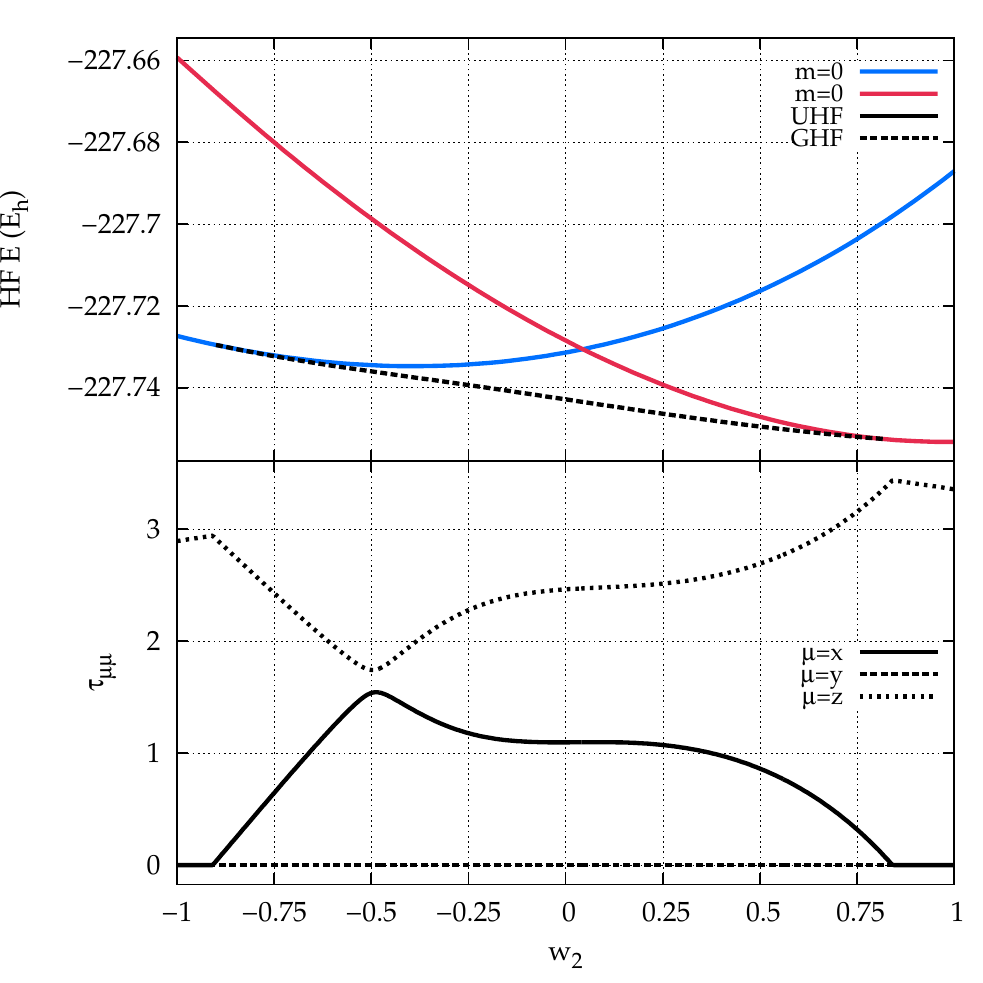}
    \includegraphics[width=0.48\textwidth]{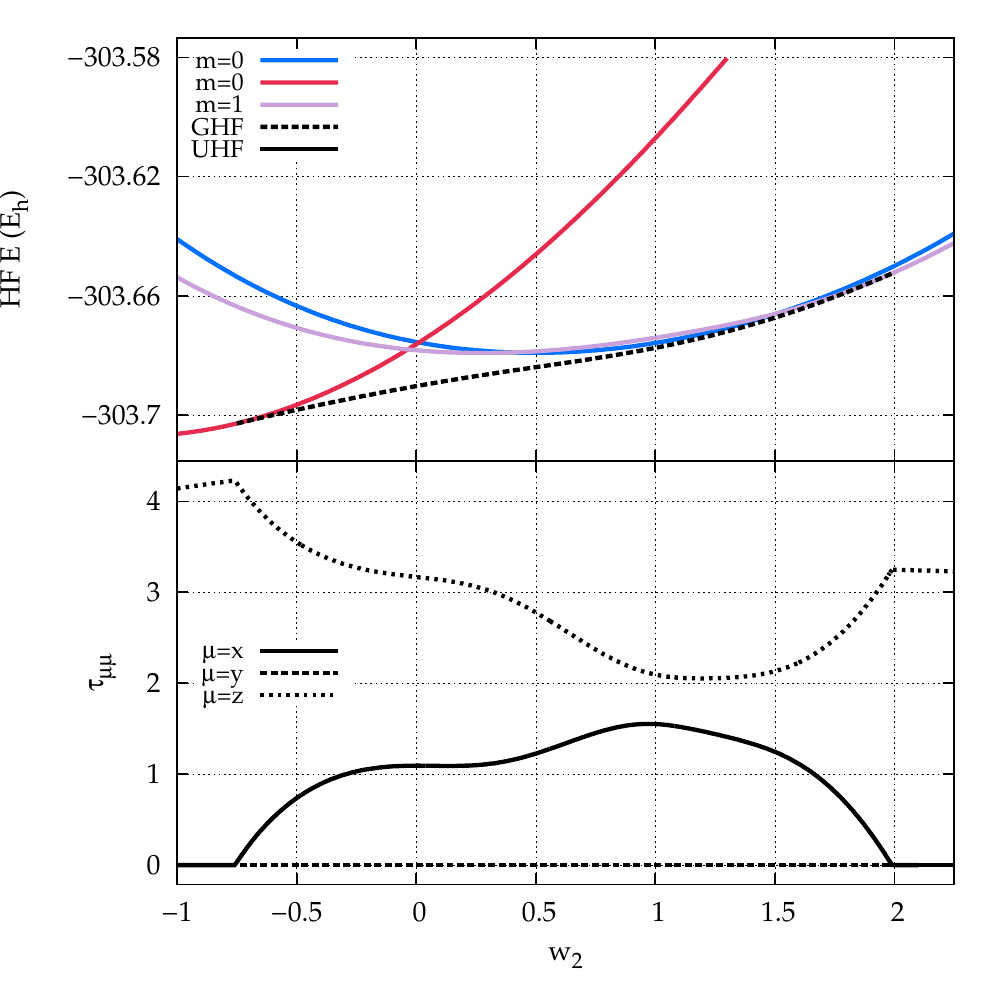}
\caption{HF energy and eigenvalues of $\mathbf{\tau}$ (Eq. \ref{coptest}) for displacement of the CX by the DC vector $\vec{x}_2$, plotted as a function of displacement weight $w_2$. Left panel: benzene. Right panel: styrene.\label{fig:benzeneandstyrene}}
\end{figure*}

Exploring Hartree-Fock along this vector reveals intersecting UHF singlets above a complex coplanar GHF ground state in each case. In benzene, the GHF solution interpolates between the UHF singlets, while in styrene the GHF solution interpolates between one of the UHF singlets and a UHF triplet that crosses the singlets nearby (Fig. \ref{fig:benzeneandstyrene}). Attempts to follow this UHF singlet fail after a sudden change in spin properties, as seen in in the vanishing solutions of cyclobutadiene.

The CX geometry in fulvene (Fig. \ref{fig:other3cx}) is achieved by twisting the methylene group approximately 30$^\circ$ \cite{CXinspec_Yarkony, fulveneCX} from the planar ground state geometry. Displacement along the GD vector $\vec{x}_1$ results in pulling the two carbons opposite the ring's substituent close together and pulling the methylene group away from the ring. UHF singlets intersect near the CASSCF CX, one of which fails to converge for larger positive weights. A complex coplanar GHF interpolates between these, though along the other branching plane vector the solution never joins UHF and remains noncollinear.

\subsection{T$_\textrm{d}$ H$_8^{+2}$}
\begin{figure}[b]
    \includegraphics[width=0.48\textwidth]{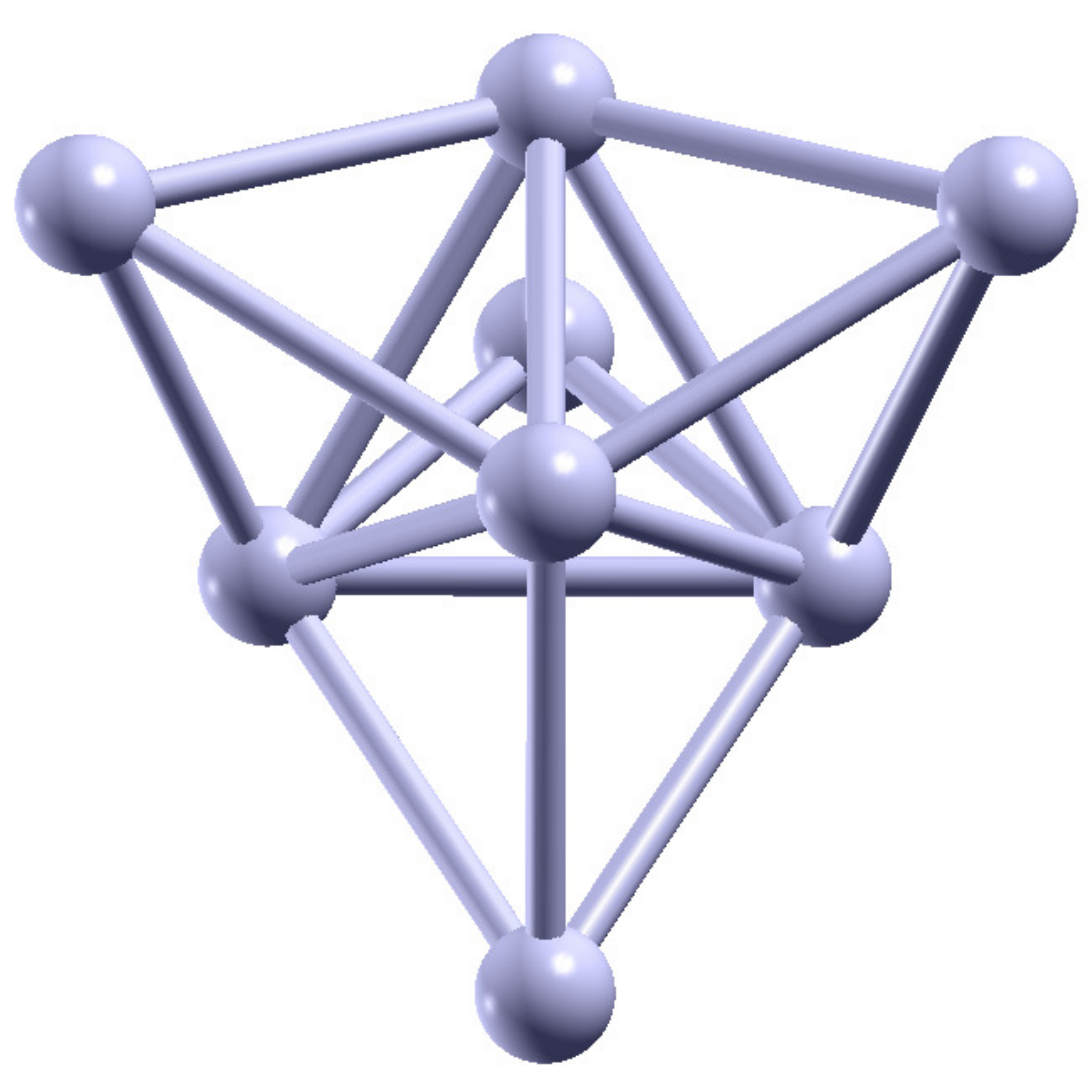}
\caption{Geometry of tetrahedral H$_8^{2+}$.
\label{fig:H8geom}}
\end{figure}

While the GHF solutions found in the branching planes discussed here break all symmetries of the Hamiltonian, spin remains coplanar in all cases. To observe noncoplanar spin, we turn to the Jahn-Teller active H$_8^{+2}$ (Fig. \ref{fig:H8geom}). Here, we have taken the tetrahedral H$_4$ model and decorated each surface of the tetrahedron with an additional hydrogen atom, resulting in a structure that is also tetrahedral. Examining the MO structure of the lowest energy real RHF solution, the neutral species has a triply degenerate HOMO due to point group symmetry. Removing two electrons results in a ground state degeneracy that is eliminated upon distortion to lower symmetry point groups. 

Thus, there is a Jahn-Teller mandated CX in this tetrahedral H$_8^{2+}$ for every H-H bond length.  Unlike our previous examples, the stable solution here is complex and noncoplanar. The high symmetry leads to a density matrix structured such that $\mathbf{M}_x = \mathbf{M}_y = \mathbf{M}_z$.  This is reflected in the eigenvalues of $\bm{\tau}$ for a H-H bond length of 1.67 \AA,  seen in Tab. \ref{table:H8T}. While removing just one electron would also result in a Jahn-Teller active ion, for H$_8^{+}$ the stable solution is collinear. It seems that having a different number of $\uparrow$- and $\downarrow$-spin electrons interferes with the spin frustration introduced by the tetrahedral geometry.

\begin{table}
\caption{Eigenvalues of $\bm{\tau}$ for the stable GHF of H$_8^{+2}$.
\label{table:H8T}}
\begin{tabular}{cc}
\hline\hline
Element      & Value \\
\hline
$\tau_{xx}$   & 1.147 \\
$\tau_{yy}$   & 1.147 \\
$\tau_{zz}$   & 1.147 \\
\hline\hline
\end{tabular}
\end{table}

\section{Discussion}
In the branching planes of conical intersections we observe HF solutions that break all symmetries, including those not represented by quantum numbers ($\hat{K}$ and $\hat{\Theta}$). Use of a recently developed magnetization diagnostic revealed that all GHF solutions found in these branching planes are coplanar. The same diagnostic identified a noncoplanar GHF solution in the Jahn-Teller active tetrahedral geometry of H$_8^{+2}$. It seems that while the spin frustration introduced by this highly symmetric geometry will lead to noncoplanar spin in HF, the strong correlation around a CX will not.  Our work here suggests that we will need to deliberately break and projectively restore both $\hat{S}^2$ and $\hat{S}_z$ symmetries as well as point group and complex conjugation or time reversal.

While the spontaneous symmetry breaking seen here precludes the use of Hartree-Fock in the description of conical intersections, it is encouraging for projected Hartree-Fock methods.  Even the simple projection after variation formalism will restore good symmetries, and should allow for the description of conical intersections reasonably well.  Even better is to use the variation after projection approach, in which the mean-field determinant is optimized in the presence of the symmetry projector rather than in its absence.  Either way, symmetry projection by means of a NOCI will lead to multireference wave functions obtained with loosely mean-field computational cost in the vicinity of a CX.  This seems a logical consequence of the breakdown of the Born-Oppenheimer approximation and our mean-field attempt to describe dynamics on multiple potential energy surfaces.

While the present results show some of the qualitative features of the CASSCF branching plane reflected in HF potential energy surfaces, there are some inconsistencies. Namely, it seems our CASSCF definition of the branching plane in cyclobutadiene only lifts HF degeneracy along one of its defining vectors.  This suggests but does not guarantee that we will see a different branching plane at the PHF level.  If symmetry breaking and restoration is to be considered an affordable alternative to the CASSCF or Full Configuration Interaction (FCI) levels of theory, a necessary step is to confirm that HF excited states will exhibit the same phenomena, such as CXs, that we are able to observe these higher levels of theory. Throughout, we have been cautious in the language used to describe HF degeneracies. 

As our symmetry broken solutions do not have good quantum numbers, they cannot define a CX or CX seam. The symmetry restored solutions, however, could potentially be optimized to CX geometry. Characteristic of a CX is the appearance of an observable geometric phase known as the Berry phase, whose existence is reliant on preservation of time reversal $\hat{\Theta}$. This effect is not exclusive to conical intersections; it emerges in any situation where there is coupling to variables, in this case  nuclear degrees of freedom, that have been excluded from the Hilbert space of the eigenvalue problem. It is nonlocal and can be observed in any wavefunction that traverses a closed loop containing the CX \cite{Berryphase, Resta_Berryreview,batista}. Projection of $\hat{\Theta}$ and calculation of this observable for a PHF CX provides an interesting future direction of this work. 

There is clearly work to be done, but the current results are promising: it may be possible to tap into the potential of symmetry breaking and restoration in HF as an FCI alternative with mean-field computational scaling.

\begin{acknowledgments}
This work was supported by the National Science Foundation under Award No. CHE-1462434. G.E.S. is a Welch Foundation Chair (No. C-0036). We would like to thank Irek Bulik and Yao Cui for development of \textit{Gaussian} links for GHF Hessian diagonalization, and further thank Irek Bulik for development of an additional \textit{Gaussian} link for testing the collinearity of GHF solutions.
\end{acknowledgments}

\bibliography{main}

\end{document}